\documentclass[aps,twocolumn,nobalancelastpage,amsmath,amssymb,
    nofootinbib,colorlinks=true,citecolor=blue,urlcolor=blue]{revtex4}
\usepackage[utf8]{inputenc}
\usepackage{float}
\usepackage{amsmath }
\usepackage{pifont}
\newcommand{\cmark}{\ding{51}}%
\newcommand{\xmark}{\ding{55}}
\usepackage{tikz}
\usepackage{graphicx}
\usepackage{dcolumn}
\usepackage{bm}
\newcommand{\bZ}{\mathbb{Z}}
\newcommand{\bG}{\mathbb{G}}
\newcommand{\cO}{\mathcal{O}}
\usepackage{hyperref}
\hypersetup{
    pdftitle={TQFT and SSB},
    unicode=false,          
    pdftoolbar=true,        
    pdfmenubar=true,        
    pdffitwindow=false,     
    pdfstartview={FitH},    
    pdfsubject={},   
    pdfcreator={},   
    pdfproducer={}, 
    pdfkeywords={} {} {}, 
    pdfnewwindow=true,      
    colorlinks=true,       
    linkcolor=magenta, 
    citecolor=blue,        
    filecolor=magenta,      
    urlcolor=blue           
} 
\begin{document}
\title{TQFT, Symmetry Breaking, and Finite Gauge Theory in 3+1D}
\begin{abstract}
We derive a canonical form for 2-group gauge theory in 3+1D which shows they are either equivalent to Dijkgraaf-Witten theory or to the so-called ``EF1" topological order of Lan-Wen. According to that classification, recently argued from a different point of view by Johnson-Freyd, this amounts to a very large class of all 3+1D TQFTs. We use this canonical form to compute all possible anomalies of 2-group gauge theory which may occur without spontaneous symmetry breaking, providing a converse of the recent symmetry-enforced-gaplessness constraints of C\'ordova-Ohmori and also uncovering some possible new examples. On the other hand, in cases where the anomaly is matched by a TQFT, we try to provide the simplest possible such TQFT. For example, with anomalies involving time reversal, $\mathbb{Z}_2$ gauge theory almost always works.
\end{abstract}
\author{Ryan Thorngren}
\affiliation{Center of Mathematical Sciences and Applications, Harvard University, Cambridge, MA 02138}
\date{\today}
\maketitle

\section{Introduction}

In recent years there has been much activity using anomaly-matching to probe the infrared physics of gauge theories \cite{Gaiotto_2017,Gomis_2018,GUO2018244,wan2019standard}. Of particular interest is the role of 1-form symmetries, whose spontaneous breaking implies deconfined gauge degrees of freedom in the IR. In the presence of a nontrivial 't Hooft anomaly, spontaneous symmetry breaking (SSB) is a typical outcome. The question arises: can we match an anomalous 1-form symmetry with a gapped phase, ie. a TQFT, without SSB? What about more general combinations of 0-form, 1-form, and gravitational anomalies? Or time reversal symmetry?

These are important questions also for probing the phase diagram of lattice systems with a Lieb-Schultz-Mattis (LSM) constraint \cite{LIEB1961407,Oshikawa_2000,Hastings_2004}, which implies a 't Hooft anomaly in the IR \cite{Cho_2017,Metlitski_2018}. Such theorems have been used to search for candidate spin liquids \cite{Savary_2016} by attempting to rule out SSB states such as magnetic order. Our method produces the simplest possible gapped spin liquid states in 3+1D consistent with a given LSM anomaly (which may be computed as a group cohomology class using the methods of \cite{else2019topological}), when there is no symmetry breaking.

Perturbative, or local, anomalies such as the chiral anomaly give nontrivial constraints on the correlation functions of local operators, so these cannot be matched by a gapped phase (and SSB implies the existence of gapless Goldstone modes). For global anomalies on the other hand, it seems some may be matched by a gapped phase without SSB, while others cannot. Recently some interesting general constraints on the anomaly have been derived assuming only topological invariance and the lack of SSB \cite{cordova2019anomaly,cordova2019anomaly2}. The general phenomenon of an anomaly which is not realizable by any gapped phase without SSB we refer to as ``symmetry protected gaplessness".

In this note, we start with 3+1D Crane-Yetter/2-group gauge theory\footnote{We do not need to consider 3-group gauge theory, because the 3-form field can always be dualized to a local order parameter. That is, 3-group gauge theory always describes a spontaneous symmetry breaking phase.} and compute all possible anomalies which can be realized without SSB, using the group cohomology and cobordism classification of anomalies \cite{CGLW,kapustin2014symmetry}. The result of our calculations are consistent with the constraints of \cite{cordova2019anomaly,cordova2019anomaly2} and in most cases we find a converse to their results---that is, all anomalies not ruled out by \cite{cordova2019anomaly,cordova2019anomaly2} are realized by a 2-group gauge theory without SSB. There is only one case where an anomaly is missing from known 3+1D TQFTs, but is not known to be ruled out by \cite{cordova2019anomaly,cordova2019anomaly2}, which we discuss in Section \ref{subsubsecpureoneform}.

It has also been argued in \cite{Lan_2019} that bosonic unitary 3+1D TQFTs are highly constrained, admitting a certain canonical gapped boundary condition where all bosonic quasiparticles are confined. To facilitate the calculation of the anomaly, we show a reduction of a general 2-group gauge theory to a canonical form, essentially a 1-form gauge theory, which matches this conjecture, and generalizes the dualities in \cite{highersymm}. We find that 2-group gauge theories realize all ``EF1" topological orders, according to the notation of \cite{Lan_2019}. There are some known topological orders which are outside this class, so until it is known how to compute anomalies of these more general theories, we cannot yet give a full computational rederivation of \cite{cordova2019anomaly,cordova2019anomaly2}, even assuming the conclusions of \cite{Lan_2019}. However, this larger class ``EF2" differs from EF1 only by certain $\bZ_2$ extensions \cite{Cui_2019}, while our ``missing" anomalies are typically of odd order, so we expect these anomalies are also not realized by EF2 topological orders.

Throughout, whenever possible, we attempt to construct the simplest gapped realization of each anomaly. For instance, most time reversal anomalies are realized by $\bZ_2$ gauge theory (aka the 3d Toric Code), the simplest 3+1D TQFT. Recently a mathematical justification for Lan-Wen's conjectured classification was given in \cite{johnsonfreyd2020classification}. This suggests that the TQFTs we find in this note are indeed the simplest possible which can match each anomaly.

We summarize our anomaly calculations in Table I.

\begin{table*}\label{table}
\centering
\caption{Global Anomalies Realized by 3+1D Finite Gauge Theory}
 \begin{tabular}{||c c c c c ||}
 \hline
 $(i,j,k)$ in classification \eqref{eqnclassanom} & Type & All Realized? & Section & Comments \\
 \hline\hline
 (6,0,-1) & pure 0-form & \xmark & \ref{subsubsecpurezeroform} & mixed finite/connected terms not realized \\ 
 \hline
 (3,3,-1) & mixed 0/1-form & \cmark & \ref{subsubsecmixedanom}  &\\
 \hline
  (1,5,-1) & mixed 0/1-form & \xmark & \ref{subsubsecmixedanom2}  & $\frac{1}{2} A \cup B \cup B$ realized \\
 \hline
 (0,6,-1) & pure 1-form & \xmark & \ref{subsubsecpureoneform}  & $(1/4) B \cup dB$ realized\\
 \hline
 (2,0,3) & mixed 0-form/gravity & \xmark & \ref{subsubsecmixedgrav} & $(1/2) A \cup w_2(TX) \cup w_2(TX)$ realized \\
 \hline
 (0,0,5) & pure gravity & \cmark & \ref{subsubsecpuregrav} & \\
 \hline\hline
  $(i,j,k)$ in classification \eqref{eqnclassanomtrs} & Type & All Realized? &  & Comments \\
 \hline\hline  (0,0,5) & pure time reversal symmetry & \cmark & & realized by $\bZ_2$ gauge theory \\ \hline
 (2,0,3) & TRS/0-form & \cmark & & realized by $\bZ_2$ gauge theory \\  \hline
 (0,2,3) & TRS/1-form & \cmark & & realized by $\bZ_2$ gauge theory \\ \hline
 (3,0,2) & TRS/0-form & \cmark & & realized by $\bZ_2$ gauge theory \\ \hline
 (1,2,2) & TRS/0-form/1-form & \cmark & & realized by $\bZ_2$ gauge theory \\ \hline
 (0,3,2) & TRS/1-form & \cmark & & realized by $\bZ_2$ gauge theory \\ \hline
 (2,2,1) & TRS/0-form/1-form & \cmark & & realized by $\bZ_2$ gauge theory \\ \hline
 (4,0,1) & TRS/0-form & \cmark & &  \\ \hline
 (0,4,1) & TRS/1-form & \xmark & & none are realized \\
  \hline
\end{tabular}
\end{table*}

\section{Symmetry Breaking and Fractionalization}\label{secbreakingandfrac}

Ordinary global symmetries in field theory of $d+1$ spacetime dimensions are associated with extended topological operators of dimension $d$. Such an operator inserted along a spacial slice indicates an application of the global symmetry operator on the Hilbert space associated with the slice, while an insertion tranverse to the slice introduces symmetry-twisted boundary conditions for the fields describing that Hilbert space.

The program of higher symmetry is to understand the symmetry principles behind general topological operators, including ones of smaller dimension \cite{highersymm,gaiotto2017symmetry} and without inverse \cite{Chang_2019,Ji_2019,thorngren2019fusion}. A \emph{$k$-form symmetry} is by definition associated with a topological operator of dimension $d-k$. They are so called because in the case of a continuous symmetry, the global parameter is a (closed) differential $k$-form.

For example, a typical 1-form symmetry is a symmetry of a gauge theory which acts by shifting the gauge field by a closed 1-form, or more generally a flat connection. In adjoint QCD, if this flat connection has holonomy in the center of the gauge group, it defines a symmetry. For some interesting consequences of this fact, see \cite{Gaiotto_2017}.

Most generally, a $k$-form symmetry associated with a topological operator $S$ acts on all operators $\cO$ of dimension $\ge k$ by wrapping $\cO$ with $S$. We say that this symmetry is spontaneously broken if there is a $k$-dimensional operator $\cO$ which is $S$-charged and has \emph{long range order}, in the sense that $\langle \cO \rangle$ decays as the area of $\cO$ rather than the volume of a region filling it in \cite{lake2018higherform}. For $k = 0$, the ordinary symmetry case, this says that $\langle \cO \rangle \neq 0$, so $\cO$ is an order parameter implying a ground state degeneracy on a sphere. For $k = 1$, this says that $\langle \cO \rangle$ obeys the perimeter law, which is the usual Wilson-'t Hooft criterion for confinement \cite{gukov2013topological}, ie. the broken phase is the one where $\cO$ is a deconfined line operator.

Another important concept is \emph{symmetry fractionalization}. This occurs when the junctions of topological operators act nontrivially on some observables \cite{Barkeshli_2019}. For instance, ordinary symmetry operators corresponding to group elements $g_1$, $g_2$, $g_1 g_2 \in G$ form a three-way junction of dimension $d-1$. A line operator may have a nontrivial linking phase with this junction. When it does, it means that line operator ferries a particle with a projective (or fractional) $G$ charge. Symmetry fractionalization is key for TQFTs to have nontrivial $0$-form anomalies \cite{anomaliesthree,anomaliesvarious}.

\begin{figure}
    \centering
    \includegraphics[width=0.8\columnwidth]{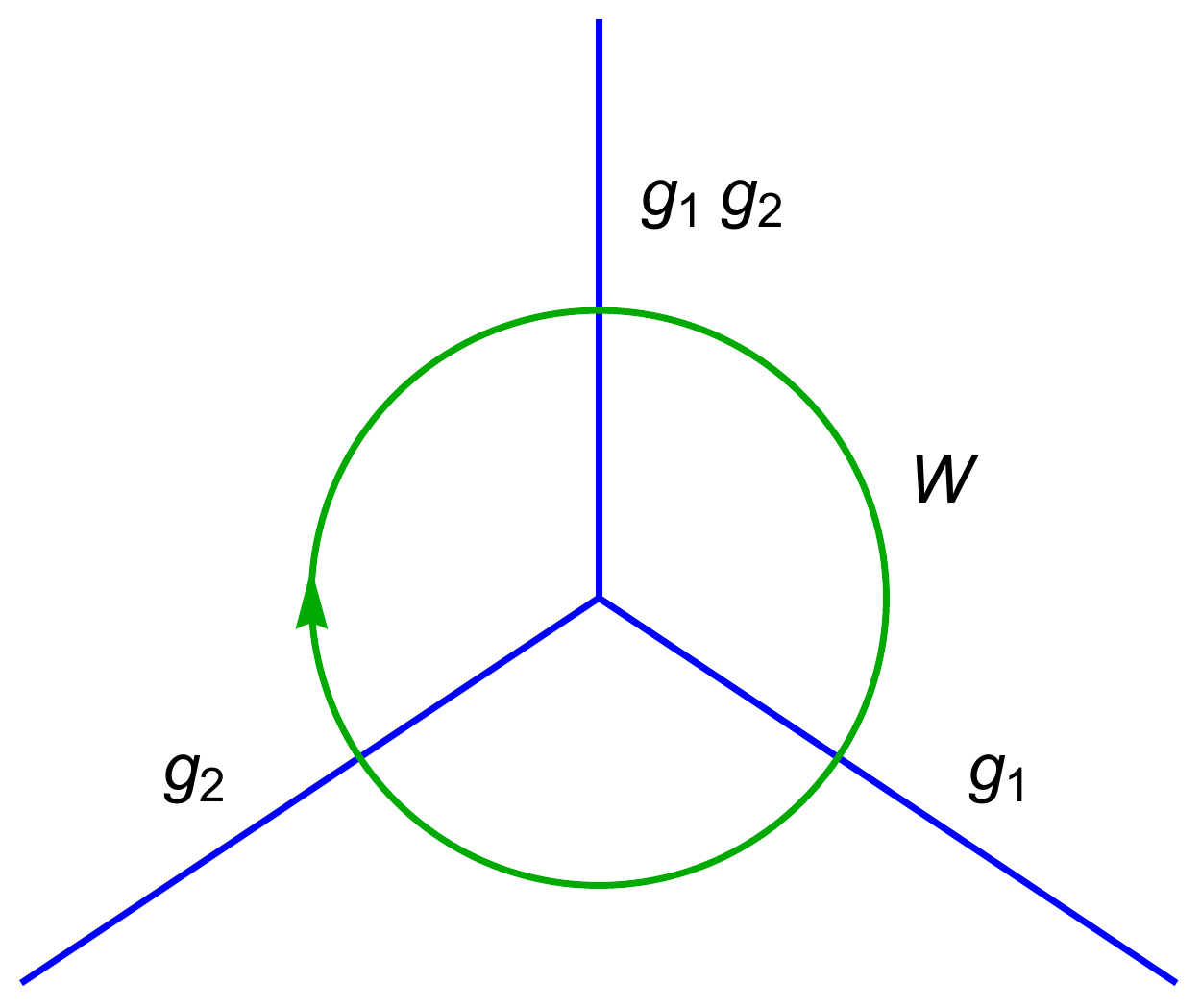}
    \caption{A typical symmetry fractionalization pattern where a 3-fold junction of 0-form symmetry defects corresponding to the group elements $g_1,g_2,g_1g_2$ (blue lines) acts on a line operator $W$. In the case that $W$ is a Wilson line, we understand this in terms of the global symmetry being nontrivially extended by the gauge symmetry. In some situations, the line $W$ changes type as it passes through the symmetry defects. This complicates the description of symmetry fractionalization \cite{Barkeshli_2019,Etingof_2010}. However, in 3+1D gauge theory, such ``anyon-permuting" symmetries are highly constrained, and amount to an action of the global symmetry on the gauge symmetry by group automorphisms.}
    \label{fig:my_label}
\end{figure}

A simple way to encode symmetry fractionalization is to think in terms of SPTs. Indeed, a $k$-form symmetry is fractionalized on some extended $l$-dimensional object, $l > k$, if that object carries a $l$-dimensional SPT for that symmetry. In the case $l = k$, $k$-dimensional SPT classes are just symmetry charges, so this is the familiar data of the symmetry action.

The assignments of these SPTs are constrained by the fusion rules of the objects. This can lead to a simplification of the symmetry fractionalization. For example, in finite $G$ gauge theory in space dimension $d \ge 2$, a 0-form symmetry $K$ may be fractionalized on codimension 2 (spacetime dimension $d-1$) 't Hooft operators (aka gauge fluxes). The data may be summarized by a single class in $\alpha \in H^{d-1}(BK,G^*)$, where $G^* = {\rm Hom}(G,U(1))$\footnote{We include cases where $K$ acts nontrivially on $G^*$, in which case these cohomology groups are understood using twisted coefficients. This data is exhaustive for $d \ge 3$, but for $d = 2$ the symmetry can mix Wilson and 't Hooft operators. A full description requires the machinery of \cite{Barkeshli_2019,Etingof_2010} and leads to some very interesting realization of anomalies, see eg. \cite{delmastro2019symmetries}.}. A 't Hooft operator corresponding to an element $g \in G$ carries the $K$-SPT $\alpha(g) \in H^{d-1}(BK,U(1))$. This may be straightforwardly modified to include ``beyond cohomology" symmetry fractionalization, as we will also consider below.

We will see how these symmetry fractionalization classes are captured by non-minimal coupling terms in the gauged action---terms which are higher order in the background gauge field than the usual $j^\mu A_\mu$.

Nontrivial $d-1$-form symmetries of TQFTs in $d+1$ dimensions are always spontaneously broken. The reason is that the line operators which generate these symmetries cannot fractionalize, since their junctions are local operators in spacetime, and TQFTs have no nontrivial local operators. (For the same reason, $0$-form symmetries of TQFTs are always unbroken.)

In 2+1D, this follows from the results of \cite{Etingof_2010}, which imply that nontrivial 1-form symmetries are in one-to-one correspondence with the abelian anyons. By modularity, every abelian anyon has a dual anyon in the TQFT it braids with. This dual anyon is deconfined by definition, so any nontrivial 1-form symmetry is spontaneously broken. For this reason, 3+1D is the most interesting dimension to find symmetry protected gaplessness.

\section{Anomalies}\label{secanom}

In this section, we compute all possible anomalies of 3+1D finite gauge theory, the results of which are summarized in the table. We freely use the cocycle theory of discrete gauge fields, which we review in the appendix. In each case that the anomaly is realized by a TQFT, we will try to determine the ``simplest" such theory. It seems likely that every anomaly considered here has been realized somewhere in the literature, but as far as I know they have not all appeared together in the same place before.

We argue below the most general 3+1D finite gauge theory consists of a 1-form $G$ gauge field $a \in Z^1(X,G)$ ($G$ possibly nonabelian) and possibly also a $\bZ_2$ 2-form gauge field $b$ (in the presence of which we have fermionic quasiparticles, otherwise they are all bosons) which satisfies
\begin{equation}\label{eqnpostnikov1}db = \beta(a) + s w_3(TX),\end{equation}
where $\beta \in Z^3(G,\bZ_2)$ is the Postnikov class \cite{highersymm}, $w_3(TX)$ is the third Stiefel-Whitney class \cite{milnor1974characteristic}, and $s \in \bZ_2$ describes whether the $b$ Wilson string is fermionic \cite{thorngren2015framed} (in this case we have the gravitational anomaly in Section \ref{subsubsecpuregrav}). The most general action is
\begin{equation}\label{eqnEF1action}S_0 = \omega(a) + \frac{1}{2} (\gamma(a) + b) \cup b,\end{equation}
where $\omega$ is a Dijkgraaf-Witten-like topological term \cite{dijkgraafwitten}, $\gamma$ describes string-like $a$-defects which are charged under $b$. The consistency conditions are described around \eqref{eqncanongaugeinv}. In the absence of $b$, we simply have $d\omega = 0$.

\subsection{Symmetry Actions}

Global symmetry actions are captured by coupling to background gauge fields. We will use $A$ (resp. $B$) to denote a background 1-form (resp. 2-form) gauge field which couples to a 0-form $K_0$ (resp. 1-form $K_1$) global symmetry. The most basic sort of coupling is the \emph{minimal coupling} of the form $j_1 \cup A$ or $j_2 \cup B$, where $j_1$ (resp. $j_2$) is a $d$ (resp. $d-1$) cocycle, a ``discrete current" made from the dynamical fields, and represents the density of charged particles (resp. strings). Charge conservation is equivalent to $dj_1 = 0$ (resp. $dj_2 = 0$).

For example, with $K_1 = \bZ_2$ we could have the minimal coupling
\begin{equation}\frac{1}{2} B \cup b.\end{equation}
This 1-form symmetry is generated by the $b$ Wilson surfaces. It is spontaneously broken because the $b$ 't Hooft lines are deconfined. On the other hand, a coupling such as
\begin{equation}B \cup \eta(a),\end{equation}
where $\eta(a) \in H^2(BG,K_1^*)$, describes 1-form charges of string-like intersection of the $a$ domain walls. This does not imply SSB---rather it is a form of symmetry fractionalization. There are also non-minimal couplings of the form
\begin{equation}\theta(B) \cup a,\end{equation}
where $\theta \in H^3(B^2 K_1, G^*)$ describes how the 1-form symmetry is fractionalized on $a$ 't Hooft surfaces. In some cases,

We denote these sorts of symmetries as \emph{magnetic}, since magnetic operators, such as 't Hooft surfaces, are charged (or symmetry-fractionalized) while electric operators, such as Wilson lines, have trivial symmetry action. Such symmetries are always anomaly free, because the coupling terms we've added are manifestly gauge invariant under all transformations.

There are also \emph{electric} symmetries, which act on the Wilson operators. The coupling of such symmetries to the background gauge fields are by modifying the cocycle constraints of the gauge fields. For a 0-form symmetry, the general form is
\begin{equation}\begin{gathered}
d_\alpha a = \mu(A)\\db = \beta(a) + s w_3(TX) + \nu(A),\end{gathered}\end{equation}
where $\mu \in H^2(BK_0,G)$, $\nu \in H^3(BK_0,\bZ_2)$, $w_3(TX)$ is the 3rd Stiefel-Whitney class (see Section \ref{subsecgravcoupling} and elsewhere below), and $d_\alpha$ denotes the twisted differential, defined by an action $\alpha$ of $K_0$ on $G$ (all coefficients are understood with respect to this action). $\nu$ has the interpretation of a projective $K_0$ action on the $G$ Wilson lines, while $\nu(A)$ is a form of $K_0$ symmetry fractionalization on the $b$ Wilson surfaces. Neither of these couplings leads to symmetry breaking.

Meanwhile, the most general electric symmetry coupling for a 1-form symmetry is
\begin{equation}\begin{gathered}
da = g(B)\\db = \beta(a) + s w_3(TX) + \kappa(B),\end{gathered}\end{equation}
where $g:K_1 \to G$ determines the 1-form charge of $a$ Wilson lines (leading to SSB) and $\kappa \in H^3(B^2 K_1, \bZ_2)$ defines the 1-form symmetry fractionalization of $b$ Wilson surfaces.

\subsection{Classifying Anomalies}

The total group of anomaly polynomials for a product 0-form $K_0$ and 1-form $K_1$ global symmetry of a $d+1$ dimensional theory can be written in cohomology as
\begin{equation}\label{eqnclassanom}
    \left(\bigoplus_{i+j+k = d + 2} H^i(BK_0, H^j(B^2 K_1, \Omega^k_{SO}))\right) {\Big /} \sim,
\end{equation}
where $BK_0$ is the classifying space for the 1-form gauge field, $B^2 K_1$ is the classifying space for the 2-form gauge field, $\Omega^k_{SO}$ are the Anderson duals of the oriented bordism groups \cite{FreedHopkins}, which contain Stiefel-Whitney classes as well as gravitational Chern-Simons terms \cite{kapustin2014symmetry,FSPT}, and the quotient indicates identification of classes by the Wu formulas. Note we do not assume finite $K_1$ or $K_0$, although the former must be abelian, and we do assume compactness. The relevant groups for us are
\begin{equation}
    \begin{gathered}
    \Omega^{-1}_{SO} = \bZ \\
    \Omega^2_{SO} = \bZ_2^{w_2} \\
    \Omega^3_{SO} = \bZ^{p_1} \\
    \Omega^5_{SO} = \bZ_2^{w_2 w_3}.
    \end{gathered}
\end{equation}
In Appendix \ref{appcomptheclass}, we describe how to use this data to compute \eqref{eqnclassanom}. The $k = -1$ piece corresponds to non-gravitational anomalies. The $k = 2$ piece always reduces to a mod 2 $k = -1$ term by a Wu formula. Each factor in \eqref{eqnclassanom} is zero or can be reduced to one of the six $(i,j,k)$'s listed in the table.

\subsection{Realizing Anomalies}

\subsubsection{Pure 0-form Anomalies}\label{subsubsecpurezeroform}

0-form symmetries are always unbroken in finite gauge theory, since it lacks local operators. 0-form anomalies of finite gauge theories has been well studied \cite{anomaliesthree,anomaliesvarious,thorngren2015higher,tachikawa2017gauging,WangWenWitten}. Let us review some of the results.

There are two ways of realizing pure 0-form anomalies in $G$ gauge theory. One is to mix magnetic and electric couplings, eg.
\begin{equation}
\begin{gathered}
a \cup \zeta(A) \\
da = \mu(A),
\end{gathered}
\end{equation}
where $\zeta \in H^3(BK_0,G^*)$ describes symmetry fractionalization on 't Hooft surfaces and $\mu \in H^2(BK_0,G)$ describes symmetry fractionalization on Wilson lines. In the absence of extra couplings or pure topological terms for $a$, the anomaly is simply computed by taking the differential of the first coupling above, using the second coupling (see \cite{anomaliesvarious} for explanations of this fact). It is
\begin{equation}\mu(A) \cup \zeta(A).\end{equation}
Any anomaly polynomial which may be decomposed as such a product can be realized this way.

Another way is to have just the second coupling above, but to also have a pure topological term for $a$. In this case, $\mu$ defines a (possibly non-central) group extension
\begin{equation}\label{eqnextension}G \to \hat G \to K_0\end{equation}
and to find the anomaly we must study the extension problem for this topological term to $\hat G$ \cite{anomaliesthree}. This was done in \cite{anomaliesvarious} using the Serre spectral sequence (see also \cite{thorngren2015framed,tachikawa2017gauging,WangWenWitten}). The results of \cite{tachikawa2017gauging} (Section 2.7) imply that for \emph{finite} $K_0$, we can find such an extension and a $G$ topological term which realizes the anomaly. Note that certain $K_0$ (such as the exceptional binary polyhedral groups) lack any nontrivial central extensions. In these cases it is necessary that $K_0$ act on $G$, ie. be an anyon-permuting symmetry, to realize the anomaly.\footnote{Note that the spectral sequence of the group extension \eqref{eqnextension} can have nontrivial differentials even in the semidirect product case, that is, without symmetry fractionalization! See \cite{totaro1996cohomology} for examples. It would be interesting to see if this happens in any physically-relevant situations.}

On the other hand, simply connected Lie groups have no central extensions and cannot act nontrivially on any finite $G$, so they are natural candidates for a global $K_0$ anomaly with symmetry protected gaplessness. \cite{borel} contains results to rule out $SU(n)$, $Sp(n)$, $Spin(n)$, $G_2$, and $F_4$. Meanwhile, the classifying spaces of $E_6$, $E_7$, and $E_8$ are well approximated by $B^4 \bZ$ in low degrees (at least up to their 8-skeleton) and can be ruled out this way \cite{henriquesMO}.

Non-simply connected (but still connected) Lie groups do have global anomalies in 3+1D (such as $\frac{1}{2} w_2 w_3$ for $SO(n)$), but by taking the extension \eqref{eqnextension} so that $\hat G$ is the universal cover of $K_0$ (eg. $Spin(n)$), by the above analysis in the simply connected case we can always match the anomaly with just symmetry fractionalization.

However, certain mixed anomalies between simply connected Lie groups and finite groups appear to have symmetry-protected gaplessness. Indeed, like $G$ be a nonabelian simply connected compact Lie group. Then $H^4(BG,\bZ) = \bZ$, while lower cohomology groups are zero. Let $c_2(A_{\rm cont})$ denote this generator (it generalizes the second Chern class for $G = SU(m)$). Then, there is a mixed anomaly for $G \times \bZ_n$ of the form
\begin{equation}\label{eqn0formgaples}
    \frac{1}{n} A_n \cup c_2(A_{\rm cont}),
\end{equation}
where $A_n$ is the $\bZ_n$ gauge field. Evidently this anomaly is not realized by any finite gauge theory. When the $\bZ_n$ gauge field is extended to $U(1)$, it is a local anomaly of Chern-Simons type $F_1 \wedge F_2 \wedge F_2$. When the $\bZ_n$ gauge field is thought of as a spin structure, and $G = SU(2)$ this represents Witten's global $SU(2)$ anomaly \cite{wittenSU2anom}.

The fact that this anomaly cannot be realized by any 3+1D TQFT follows from the results of C\'ordova-Ohmori \cite{cordova2019anomaly}. They showed that if the anomaly polynomial is nontrivial on any background on $S^1 \times S^2 \times S^2$, then it cannot be realized by a TQFT without SSB. Indeed, we can construct a $G$ bundle on $S^2 \times S^2$ of instanton number 1 using the ``collapse map"\footnote{This map is constructed by considering $S^2 \times S^2$ as a quotient of $D^2 \times D^2 = B^4$ along its boundary. Then the map $S^2 \times S^2 \to S^4$ is obtained by collapsing the entire boundary to a point.} $S^2 \times S^2 \to S^4$ and the fact that $\pi_3 G = \bZ$. Then, we place $A_n$ holonomy along $S^1$ to obtain a nontrivial background for \eqref{eqn0formgaples} on $S^1 \times S^2 \times S^2$.

\subsubsection{Mixed 0-form/1-form Anomalies of Signature $(3,3,-1)$}\label{subsubsecmixedanom}

Mixed anomalies between 0-form and 1-form symmetries come in two kinds, of signature $(3,3,-1)$ and $(1,5,-1)$. We first consider the former. These again split into two types, according to the decomposition $K_1 = T \times U(1)^r$, where $T$ is a finite abelian group (see Appendix \ref{appcomptheclass}). These two cases don't have any nontrivial mixing, so we can first assume $K_1 = U(1)$. A general anomaly for this group may be written
\begin{equation}B \cup \lambda(A),\end{equation}
where $\lambda \in H^3(BK_0,\bZ)$ describes line-like defects of the $K_0$ gauge field where $K_1$-charged strings are created. For compact groups, $H^3(BK_0,\bZ)$ is torsion, so there is some $n$ and some $\xi \in H^2(BK_0,\bZ_n)$ such that
\begin{equation}\lambda = \frac{1}{n} d\xi.\end{equation}
We can thus integrate the above by parts to obtain the equivalent anomaly
\begin{equation}\frac{1}{n} dB \cup \xi(A).\end{equation}
This anomaly is realized in $G = \bZ_n$ gauge theory without SSB by $\xi(A)$ fractionalizing $K_0$ on Wilson lines and $dB/n$ fractionalizing the $U(1)$ part of $K_1$ on 't Hooft surfaces via the coupling
\begin{equation}\frac{1}{n} dB \cup a.\end{equation}

Now we assume $K_1$ is finite. The general anomaly may be written
\begin{equation}B \cup \lambda(A),\end{equation}
where $\lambda(A) \in H^3(BK_0, K_1^*)$ has the same interpretation as above. It is easy to realize this anomaly in $G = K_1$ gauge theory with SSB by giving the Wilson lines $K_1$ charges and fractionalizing $K_0$ on 't Hooft surfaces according to $\lambda(A)$.

However, to realize the anomaly in $G$ gauge theory without SSB, we need to mimic the above strategy. That is, we must find some $j_2 \in H^2(BG,K_1^*)$ and have the coupling
\begin{equation}B \cup j_2(a),\end{equation}
as well as some class $\beta \in H^2(BK_0,G)$ which fractionalizes the $K_0$ symmetry on the $G$ Wilson lines, so that $d\beta = \lambda$.

This mathematical problem is the same kind as the one we studied for realizing the pure 0-form anomalies. Indeed, the arguments in Section 2.7 of \cite{tachikawa2017gauging} can be easily adapted for $K_1^*$ coefficients to find such a pair $(j_2,\beta)$, as long as $K_0$ is finite. For infinite $K_0$ we can use the fact that simply connected Lie groups have their first nonzero group cohomology group in degree 4. In either case, we can realize any mixed anomaly by this method.

\subsubsection{Mixed 0-form/1-form Anomalies of Signature $(1,5,-1)$}\label{subsubsecmixedanom2}

The other type of mixed 0-form/1-form anomalies take the form
\begin{equation}
    \frac{1}{n} f(A) \cup P(B),
\end{equation}
where $P \in H^4(B^2 K_1,U(1))$ is order $n$, and $f:K_0 \to \bZ_n$ is a homomorphism. An anomaly of this form is realized for example by $SU(2)$ adjoint QCD \cite{cordova2018candidate} and possible gapped realizations of that anomaly are discussed in \cite{Wan_2019}. Note that the continuous component of $K_1$ does not contribute to $P$, so we may assume $K_1$ is finite. In this case, $P = P_q$ is defined by a quadratic form
\begin{equation}
    q:K_1 \to U(1)
\end{equation}
using the Pontryagin square (see Section \ref{subsecgenduality}).

In cases with even torsion, even degree generators $P_q$ cannot be written as a product of two 2-cocycles, so we cannot apply the strategy of the previous section. In this case, it appears we must break the 0-form symmetry, such that domain walls between different vacua have the anomaly $P_q$ \cite{hason2019anomaly}. See \cite{seibergkapustin} for some examples of this anomaly.

Other elements are sums of terms
\begin{equation}
    P_q(B) = \frac{1}{n} B_i \cup B_j,
\end{equation}
where $B_{i,j}$ are obtained from $B$ by splitting $K_1$ into its finite cyclic factors (in the expression above, $n$ is the gcd of the orders of $B_i$ and $B_j$). These anomalies can be satisfied by just breaking the 1-form symmetry, using a $\bZ_n$ gauge theory, where the 1-form symmetry acts on Wilson lines (leading to SSB) via the coupling
\begin{equation}
    da = B_j,
\end{equation}
while we must also have mixed 0-form/1-form fractionalization on 't Hooft surfaces via the topological term
\begin{equation}
    \frac{1}{n} a \cup f(A) \cup B_i.
\end{equation}

There is one set of anomalies which may be realized without any symmetry breaking, and that is the order 2 anomaly
\begin{equation}\label{eqnmixedanomarealized}
    \frac{1}{2} f(A) \cup B \cup B
\end{equation}
where we take $K_1 = \bZ_2$ and $f:K_0 \to \bZ_2$. This anomaly is realized in a TQFT with a fermionic quasiparticle, represented by a dynamical $\bZ_2$ 2-cocycle $b$, with the action
\begin{equation}
    S = \frac{1}{2} b \cup b.
\end{equation}
We couple this theory to the background fields $A$ and $B$ such that these symmetries fractionalize on the $b$ Wilson surface:
\begin{equation}
    db = f(A) \cup B.
\end{equation}
We find after some computation (and up to adding a counterterm---see the end of Section \ref{subsecgenduality} for details)
\begin{equation}
    dS = \frac{1}{2} Sq^2 (f(A) \cup B),
\end{equation}
where $Sq^2$ is the second Steenrod square \cite{cohomologyoperations}. This expression is equivalent to \eqref{eqnmixedanomarealized}.

We can again make contact with the results of C\'ordova-Ohmori \cite{cordova2019anomaly}. Indeed, all except those of the form \eqref{eqnmixedanomarealized} anomalies have a nontrivial partition function on $S^1 \times S^2 \times S^2$, where we place the $A$ background around the $S^1$, $B_i$ around the first $S^2$, and $B_j$ around the second. In the case of an even degree anomaly which does not factorize as a product, a diagonal background $B$ on $S^2 \times S^2$ of highest possible even degree will do. The reason \eqref{eqnmixedanomarealized} does not have a nontrivial partition function on $S^1 \times S^2 \times S^2$ is because the cup square of any 2-cocycle on $S^2 \times S^2$ is always even.

\subsubsection{Pure 1-form Anomalies}\label{subsubsecpureoneform}

As we have discussed in Section \ref{subsubsecmixedanom}, the only non-SSB coupling of $G$ gauge theory to a 1-form symmetry is by the term
\begin{equation}B \cup j_2(A).\end{equation}
Thus, $G$ gauge theory has no pure 1-form anomalies without SSB.

However, if we have a fermionic quasiparticle, as in \eqref{eqnEF1action}, then there are certain mod 2 pure 1-form anomalies, as follows (these are analogous to those found in the previous section). We take $G = 1$ so there is only a dynamical $\bZ_2$ 2-form field $b$, with the action
\begin{equation}S = \frac{1}{2} b \cup b.\end{equation}
We couple the theory to a background 2-form $B$ by
\begin{equation}db = \theta(B),\end{equation}
where $\theta \in H^3(B^2 K_1, \bZ_2)$ describes how the 1-form symmetry fractionalizes on $b$ Wilson surfaces. We find
\begin{equation}dS = \frac{1}{2} Sq^2 \theta(B),\end{equation}
where $Sq^2$ is the second Steenrod square \cite{cohomologyoperations} (compare the previous section and see the end of Section \ref{subsecgenduality} for details). For example, if $K_1 = \bZ_2$ and $\theta(B) = dB/2$, then by the Adem relations we have the anomaly\footnote{Appendix C.1 of \cite{clement} contains the relevant generator in degree 6 and other useful calculations.}
\begin{equation}dS = \frac{1}{2} B \cup \frac{dB}{2}.\end{equation}
Using the Wu formula, this anomaly is the same as
\begin{equation}\frac{1}{2} w_2(TX) \cup \theta(B),\end{equation}
where $w_2(TX)$ is the 2nd Stiefel-Whitney class \cite{milnor1974characteristic}. It has the interpretation as a fractionalization anomaly: the $b$ 't Hooft line describes a fermionic quasiparticle, and the fractionalization $db = dB/2$ means that two $B$ 't Hooft lines fuse to a $b$ 't Hooft line, but it is impossible to fractionalize a fermion in 3+1D this way.

It is clear that coupling $b$ to a $G$ gauge field cannot produce more anomalies since we cannot modify the cocycle condition for $a$ using $B$ without breaking the 1-form symmetry by giving $a$ Wilson lines nontrivial charges.

This appears to be consistent with the results of C\'ordova-Ohmori \cite{cordova2019anomaly}, but proving no 3+1D TQFT can realize these other anomalies without SSB seems to be beyond their methods. To see this, we take $K_1 = \bZ_n$, without loss of generality, so $\theta(B) = k dB/n$ for some $k$. We must study backgrounds $B$ on mapping tori of the form $(S^1 \times S^3) \rtimes_f S^1$ or $(S^2 \times S^2) \rtimes_f S^1$, where $f$ is a diffeomorphism of $S^1 \times S^3$ or $S^2 \times S^2$, respectively. The former case cannot support a nontrivial-enough background to define a constraint since $S^3$ is 2-connected. In the latter case, simply taking a product is not enough because in this case we always have $\theta(B) = 0$---we need to choose a nontrivial diffeomorphism $f$. Then, to study backgrounds $B$ on the mapping torus, we use the Serre spectral sequence. We find that $H^2((S^2 \times S^2) \rtimes_f S^1, K_1)$ are in correspondence with $f$-invariant 2-cocycles on $S^2 \times S^2$.

It appears the mapping class group of $S^2 \times S^2$ is not known (although see \cite{Gabai_2019} for some recent progress in this direction). However, we can certainly cook up some elements. One is the diffeomorphism which acts antipodally on each $S^2$. A $\bZ_n$ 2-cocycle is $f$-invariant iff its exponentiated integrals are $\pm 1$ over each $S^2$. We find that if $B$ has integral $n/2$ over one of the $S^2$'s, then $dB/n$ is Poincar\'e dual to $n/2$ times that $S^2$. This means that the anomaly polynomials are all trivial on these backgrounds and we obtain no constraints.

Another is the swapping diffeomorphism which exchanges the two spheres. A $\bZ_n$ 2-cocycle $B$ is $f$-invariant if it has the same integral over each sphere. In this case, it admits an $f$-invariant integer lift, so $\theta(B) = 0$ and we again obtain no constraints.

The lack of understanding of the mapping class group of $S^2 \times S^2$ is a significant obstruction to applying the framework of \cite{cordova2019anomaly}. Perhaps there is a diffeomorphism which forbids symmetry-preserving gapped phases with pure 1-form anomalies of degree $>2$ (although I find this doubtful in view of \cite{Gabai_2019}), perhaps this symmetry-protected gaplessness can be shown by other methods, or perhaps there is even some yet-unknown TQFT which can realize this anomaly without SSB. We leave this interesting question to future work.

\subsubsection{Mixed 0-form/gravity Anomalies}\label{subsubsecmixedgrav}

There are several ways to couple $G$ gauge theory to gravity. We give a full treatment in Section \ref{subsecgravcoupling}. We have noted above that mixed 0-form/gravity anomalies of signature $(3,0,2)$ are equivalent to pure 0-form anomalies by the Wu formula. It thus suffices to study those of signature $(2,0,3)$. These are Chern-Simons-like terms descending from the integer 6-cocycle
\begin{equation}\label{eqnmixedgravanom}
  \alpha(A) \cup p_1(TX),
\end{equation}
where $p_1(TX) \in H^4(X,\bZ)$ is the first Pontryagin class and $\alpha \in H^2(BK_0,\bZ)$. In the case $K_0 = U(1)$, this becomes a mixed Chern-Simons term of type $F \wedge R \wedge R$ which implies a gapless chiral mode on the flux string, and thus cannot be realized in a gapped theory without SSB. In all other compact cases, $\alpha$ is torsion, so there exists some $n$ and a homomorphism $f:K_0 \to \bZ_n$ such that
\begin{equation}\alpha(A) = \frac{df(A)}{n}.\end{equation}
In this case \eqref{eqnmixedgravanom} may be rewritten as a 5-cocycle with $U(1)$ coefficients:
\begin{equation}
  \frac{1}{n} f(A) \cup p_1(TX).
\end{equation}
This makes it clear that the anomaly only depends on $p_1(TX)$ mod $n$. Further, by the methods of \cite{hason2019anomaly}, its form implies that if we introduce two real order parameters transforming in the $2\pi/n$ rotation representation associated to $f:K_0 \to \bZ_n$, then we can realize the anomaly in an SSB phase with a $c = 8$ chiral mode on the domain wall (the boundary mode of the gravitational Chern-Simons term with the smallest allowed level in a bosonic system).

If $n = 2$, we obtain a simplification using the formula
\begin{equation}p_1(TX) = w_2(TX) \cup w_2(TX) \mod 2.\end{equation}
Such 2-torsion anomalies may be realized in a gapped phase without SSB as follows. We need to use a theory with a fermionic quasiparticle, ie. a $\bZ_2$ 2-form $b$ with topological term as in \eqref{eqnEF1action}. Then, we want the symmetry fractionalization pattern
\begin{equation}db = f(A) \cup w_2(TX).\end{equation}
This exotic symmetry fractionalization pattern means that where the $b$ 't Hooft string intersects a $k \in K_0$ symmetry wall with $f(k) = 1$ mod 2, the intersection point, a particle-like object, is a fermion.

However, if $n > 2$, by the results of \cite{cordova2019anomaly2}, this anomaly cannot be realized by \emph{any} gapped phase (without SSB). For other discussions of realizing this and related anomalies in gapped phases, see \cite{Garc_a_Etxebarria_2019,hsieh2018discrete}.

\subsubsection{Pure Gravitational Anomalies}\label{subsubsecpuregrav}

There is one pure gravitational anomaly in 3+1D, associated with the anomaly polynomial
\begin{equation}
  \frac{1}{2} w_2(TX) \cup w_3(TX).
\end{equation}
This may be detected, for example, on the mapping torus of the complex conjugation diffeomorphism of $\mathbb{CP}^2$ \cite{Wang_2019}. It was realized by an ``all fermion" topological order in \cite{thorngren2015framed}. In particular, if we take $s = 1$ in \eqref{eqnpostnikov1}, meaning
\begin{equation}
  db = w_3(TX),
\end{equation}
which turns the $b$ Wilson string into a fermion, we find this anomaly by computing the differential of the $b\cup b$ term. See Section \ref{subsecgenduality} for details on the calculation.

\subsection{Anomalies Involving Time Reversal}

Our methods can be easily extended to the case involving time reversal symmetry (TRS). One needs only to replace the oriented bordism groups $\Omega^k_{SO}$ in \eqref{eqnclassanom} with the un-oriented bordism groups $\Omega^k_O$:
\begin{equation}\label{eqnclassanomtrs}
    \left(\bigoplus_{i+j+k = d + 2} H^i(BK_0, H^j(B^2 K_1, \Omega^k_{O}))\right) {\Big /} \sim.
\end{equation}
The calculations actually simplify quite a bit, since these groups are all 2-torsion, generated by polynomials in the Stiefel-Whitney classes, see Appendix \ref{appcomptheclass}.

Let us just briefly summarize some results in this setting. First of all, because of the overall 2-torsion in \eqref{eqnclassanomtrs}, most of the possible topological terms factorize into a product of terms of degree $\le 3$, at least one of which is a degree 2 term made from $w_1(TX)$ and $A$ which we can use a fractionalization class for a $\bZ_2$ gauge field $a$. This allows us to use the techniques of Section \ref{subsubsecmixedanom} to construct $\bZ_2$ gauge theories with the appropriate anomaly and no SSB.

For example, there is a pure time reversal anomaly (signature $(0,0,5)$)
\begin{equation}\frac{1}{2} w_1(TX)^5,\end{equation}
which may be realized in $\bZ_2$ gauge theory by combining a topological term
\begin{equation}\label{eqntrsfrac1}\frac{1}{2} w_1(TX)^3 \cup a,\end{equation}
which fractionalizes TRS on 't Hooft surfaces, with a fractionalization
\begin{equation}\label{eqntrsfrac2kram}da = w_1(TX)^2,\end{equation}
which makes the Wilson line carry a Kramers doublet \cite{thorngren2015framed}.

Another class, those of signature $(2,0,3)$, are of the form
\begin{equation}\frac{1}{2} c(A) \cup w_1(TX)^3,\end{equation}
where $c \in H^2(BK_0,\bZ_2)$. These can be realized in $\bZ_2$ gauge theory by fractionalizing $K_0$ on Wilson lines via
\begin{equation}
    da = c(A)
\end{equation}
and TRS on 't Hooft surfaces via \eqref{eqntrsfrac1}. Alternatively, we can use \eqref{eqntrsfrac2kram} so that the Wilson lines are Kramers doublets, and add the topological term
\[\frac{1}{2} a \cup w_1(TX) \cup c(A),\]
which describes a kind of $K_0 \times T$ fractionalization on 't Hooft surfaces.


There are essentially only two interesting cases: signatures $(4,0,1)$ and $(0,4,1)$. Let us first consider the former. These mixed TRS/0-form anomalies may be written
\begin{equation}
    \frac{1}{2} w_1(TX) \cup P(A),
\end{equation}
where $P \in H^4(BK_0,\bZ_2)$. By the Wu formula, any of these where $P(A)$ has an integer lift, eg. terms depending only on the continuous part of $K_0$, should be considered trivial. That leaves the continuous part of $K_0$ to enter through mixed terms with the discrete part. All such anomalies decompose into terms of degree $\le 3$ and so we can find a $\bZ_2$ gauge theory and a fractionalization pattern to realize this anomaly without SSB. Thus we may assume without loss of generality that $K_0$ is finite.

Then, using similar arguments as in Section \ref{subsubsecpurezeroform}, for finite $K_0$ we can realize this anomaly in some $G$ gauge theory with a topological term
\begin{equation}
    \frac{1}{2} w_1(TX) \cup Q(a),
\end{equation}
where $Q \in H^3(BG,\bZ_2)$, by choosing an appropriate extension of $K_0$ by $G$.

Now we turn to mixed TRS/1-form anomalies of signature $(0,4,1)$. They have a similar form as above:
\begin{equation}
    \frac{1}{2} w_1(TX) \cup P(B).
\end{equation}
First, if we allow TRS to be spontaneously broken, then $\frac{1}{2} P(B)$ describes the $K_1$ anomaly on the TRS domain wall \cite{hason2019anomaly,wang2019gauge}.

As above, terms only depending on the continuous part of $K_1$ are zero. Further, there are no mixed terms between different cyclic or $U(1)$ factors of $K_1$, since $H^*(B^2 U(1),A)$ begins in degree 3. Also, cyclic factors of odd order cannot contribute anything.

The only possibilities are pure or mixed terms among even cyclic factors, ie.
\begin{equation}
    \frac{1}{2} w_1(TX) \cup B_i \cup B_j,
\end{equation}
where $B_{i,j}$ are components of $B$ along two (possibly the same) cyclic factors. An anomaly of this form is realized for $K_1 = \bZ_2$ by the center symmetry of $SU(2)$ adjoint QCD at $\theta = \pi$ \cite{Gaiotto_2017}. See also \cite{Wan_2020,Wan_20192}. Note that by the Wu formula, this anomaly is only nontrivial if one of $B_{i,j}$ represents a $\bZ_2$ 1-form symmetry (as opposed to a $\bZ_{2^k}$ 1-form symmetry, $k > 1$). Without loss of generality we take $B_i$ to be this $\bZ_2$ 2-cocycle.

We can realize this anomaly in $\bZ_2$ gauge theory by partially breaking the 1-form symmetry via a coupling
\begin{equation}
    da = B_j
\end{equation}
if we also include the topological term
\begin{equation}
    \frac{1}{2} w_1(TX) \cup B_i \cup a
\end{equation}
which fractionalizes TRS and the 1-form symmetry on $a$ 't Hooft surfaces. However, there is no finite gauge theory which can realize this anomaly without SSB. 

Indeed, this anomaly polynomial has a nontrivial partition function on $(S^2 \times S^2) \rtimes_f S^1$, where $f$ acts as an antipodal map on one of the $S^2$'s. We take $B_i$ to have integral 1 around that $S^2$ (here it is important $B_i$ is not required by the group structure of $K_1$ to have a $\bZ_{2^k}$ lift, for any $k > 1$, since we will not be able to extend such a lift to the whole mapping torus) and $B_j$ to have integral 1 around the other $S^2$. See Section 3.5 of \cite{cordova2019anomaly}.

\section{Dualities of Topological Gauge Theories in 3+1D}

In \cite{highersymm}, several 2-group TQFTs in 3+1D were shown to be equivalent to ordinary Dijkgraaf-Witten theory. We will try to generalize those arguments and make contact with the conjecture of Lan and Wen \cite{Lan_2019}.

Basic mathematical definitions can be found in the appendix.

\subsection{Partition Function Duality}

First we will show that the partition function of 2-form gauge theory with gauge group $\Pi_2$ is equivalent to that of a 1-form gauge theory with gauge group $\Pi_2^*$.

Indeed, suppose we have a theory of a 2-form gauge field $B \in C^2(X,\Pi_2)$. The cocycle condition
\begin{equation}dB = 0\end{equation}
may be imposed by introducing a Lagrange multiplier field $\lambda \in C^1(X^\vee,\Pi_2^*)$, where $X^\vee$ indicates the Poincar\'e dual cellulation of $X$ and
\begin{equation}\Pi_2^* = {\rm Hom}(\Pi_2,U(1))\end{equation}
the Pontryagin dual group of $\Pi_2$. The cocycle condition is imposed by the action
\begin{equation}S(B,\lambda) = \int_X \langle dB, \lambda \rangle,\end{equation}
where the integral is a sum over all pairs of 3-simplices and dual 1-simplices, weighted by the pairing of the value of $dB$ on the 3-simplex and the value of $\lambda$ on the dual 1-simplex\footnote{If one forgoes the dual cellulation, instead using the cup product action $\lambda \cup dB$, one finds zero energy configurations with $dB \neq 0$ whose counting depends on the details of the triangulation \cite{thesis}. This is related to Chern-Simons zero modes on the lattice \cite{chen2019abelian}.}. We see that if $dB \neq 0$, there is some $\lambda$ which pairs nontrivially and so summing over $\lambda$,
\begin{equation}\sum_{\lambda \in C^1(X^\vee,\Pi_2^*)} e^{iS(B,\lambda)}= \begin{cases} 0 \qquad\qquad\qquad\quad\ dB \neq 0 \\ |C^1(X^\vee,\Pi_2^*)| \qquad dB = 0 \end{cases}.\end{equation}
Thus, by inserting this factor into any correlation function involving a sum over the 2-form gauge field $B$, we can relax the cocycle constraint and let $B \in C^2(X,\Pi_2)$ be a local degree of freedom.

The derivation of the duality now proceeds by summing over $B$, using the identity
\begin{equation}S(B,\lambda) = -\int_X \langle B, d\lambda \rangle.\end{equation}
In the absence of other terms in the action or operator insertions, we obtain the constraint
\begin{equation}d\lambda = 0,\end{equation}
hence an equivalent description of the partition function as that of a 1-form gauge theory with gauge group $\Pi_2^*$.

\subsection{Operator Duality}

The basic duality can also be defined in the presence of operator insertions. For example, a Wilson surface operator
\begin{equation}\int_\Sigma \chi(B),\end{equation}
where $\Sigma$ is a simplicial 2-cycle in $X$ and $\chi \in \Pi_2^*$ is a character of $\Pi_2$, can be rewritten using Poincar\'e duality as
\begin{equation}\int_\Sigma \chi(B) =\int_X \langle \chi\delta_\Sigma,B\rangle,\end{equation}
where $\delta_\Sigma \in C^2(X^\vee,\bZ)$, $\chi \delta_\Sigma \in C^2(X^\vee,\Pi_2^*)$. Thus, we can combine the Wilson surface operator with the action and find that integrating out $B$ yields the modified constraint
\begin{equation}d\lambda = \chi \delta_\Sigma,\end{equation}
which represents a 't Hooft surface for $\lambda$. This is the essence of electric-magnetic duality. Likewise, a 't Hooft loop for $B$ of charge $g \in \Pi_2$ is supported along a 1-cycle $\gamma$ in $X^\vee$, and may be written in terms of the modified constraint
\begin{equation}dB = g \delta_\gamma.\end{equation}
To impose this constraint using the Langrange multiplier $\lambda$ we use the modified action
\begin{equation}S = \int_X \langle\lambda dB - g\delta_\gamma, \lambda \rangle.\end{equation}
We see that the second term becomes the Wilson line insertion
\begin{equation}-\int_X \langle g\delta_\gamma, \lambda \rangle = -\int_\gamma \lambda(g),\end{equation}
where $\lambda(g) \in C^1(X^\vee,U(1))$ is obtained by pairing $\lambda \in C^1(X^\vee,\Pi_2^*)$ with $g \in \Pi_2$.

\subsection{Including a 2-form Component}

We can now extend the duality to the case of a 2-group $(\Pi_1,\Pi_2,\alpha,\beta)$. In particular, the Postnikov class $\beta(A) \in C^3(X,\Pi_2)$ modifies the cocycle constraint for $B$ to
\begin{equation}d_\alpha B = \beta(A),\end{equation}
equivalent to inserting 't Hooft lines for $B$ along 1-cycles dual to $\beta(A)$. When we include the Lagrange multiplier $\lambda \in C^1(X^\vee,\Pi_2^*)$, we obtain a coupling of $A$ to $\lambda$ by
\begin{equation}S = \int_X \langle d_\alpha B - \beta(A),\lambda \rangle,\end{equation}
where we have also included the twisted differential $d_\alpha B$. The action $\alpha$ of $\Pi_1$ on $\Pi_2$ defines a dual action of $\Pi_1$ on $\Pi_2^*$, and we have the identity
\begin{equation}\int_X \langle d_\alpha B,\lambda \rangle = -\int_X \langle B,d_\alpha \lambda \rangle.\end{equation}
Thus, in the absence of operator insertions or other terms in the action, integrating out $B$ yields the constraint
\begin{equation}d_\alpha \lambda = 0.\end{equation}
Combining with the constraint $dA = 0$, the pair $(A,\lambda)$ defines a 1-form gauge field for the semidirect product group $\Pi_2^* \rtimes \Pi_1$. The Postnikov class becomes a topological term
\begin{equation}\label{eqnpostclasstopterm}-\int_X \langle \beta(A),\lambda \rangle.\end{equation}

\subsection{Topological Terms}

A general discrete gauge field in 3+1D is specified by a 2-group $\mathbb{G} = (\Pi_1,\Pi_2,\alpha,\beta)$ as well as a twist $\omega \in H^4(B\bG,U(1))$. One can think of $\omega$ as a natural functional of the 2-group gauge field $(A,B)$ which must satisfy
\begin{equation}\label{eqncocycletopterm}d\omega(A,B) = 0\end{equation}
whenever $A,B$ satisfy the cocycle constraints above. This simple-looking equation is actually equivalent to gauge invariance of $\omega$ (up to boundary terms) under the gauge transformations above \cite{anomaliesvarious}.

The Serre spectral sequence allows us to understand the general topological terms beginning with those for the \emph{product} 2-group with trivial $\alpha$ and $\beta$, which has a K\"unneth formula. This means that topological terms for the product 2-group are classified by
\begin{equation}\label{eqnserre}\bigoplus_{j = 0}^4 H^j(B\Pi_1,H^{4-j}(B^2\Pi_2,U(1))).\end{equation}
Topological terms for the general 2-group with nontrivial $\alpha$ and $\beta$ but the same $\Pi_1, \Pi_2$ are generated by a subset of these terms which satisfy some extra consistency conditions coming from \eqref{eqncocycletopterm}, which we will see below. The group above is only nonzero for $j = 0$ (pure 2-form terms), $j = 2$ (mixed terms), and $j = 4$ (pure 1-form terms). Clearly the pure 1-form terms ($j = 4$) are innocuous for the duality. Let us consider the mixed terms ($j = 2$). These have the form
\begin{equation}\int_X B \cup \eta(A) + \cdots,\end{equation}
where $\eta(A) \in C^2(X,\Pi_2^*)$ comes from $\eta \in H^2(B\Pi_1,\Pi_2^{*,\alpha})$ (the superscript indicates the dual action of $\Pi_1$ on $\Pi_2^*$ defined by $\alpha$) and $\cdots$ are terms depending only on $A$ which ensure gauge invariance by
\begin{equation}\beta(A) \cup \eta(A) + d(\cdots) = 0.\end{equation}

We can define a dual cocycle $\eta(A)^\vee \in C^2(X^\vee,\Pi_2^*)$ such that
\begin{equation}\int_X B \cup \eta(A) = \int_X \langle B,\eta(A)^\vee\rangle.\end{equation}
We see therefore that when we sum over $B$ in favor of the Lagrange multiplier, we find the modified cocycle constraint
\begin{equation}d_\alpha \lambda = \eta(A)^\vee.\end{equation}
This implies that the gauge group for $(A,\lambda)$ is a (possibly non-central) extension
\begin{equation}\label{eqndualextension}\Pi_2^* \to \hat G \to \Pi_1\end{equation}
with extension class $\eta \in H^2(B\Pi_1,\Pi_2^{*,\alpha})$.

\subsection{Coupling To Gravity}\label{subsecgravcoupling}

There are also gravitational topological terms involving characteristics of the tangent bundle. For finite groups, the only ones that enter are the Stiefel-Whitney classes $w_j(TX) \in H^j(X,\bZ_2)$ \cite{kapustin2014symmetry}. See \cite{milnor1974characteristic, thorngren2015framed} for a review. For an oriented spacetime, $w_1(TX) = 0$, $w_3(TX) = \frac{1}{2} dw_2(TX)$. Thus, by integrating by parts, the most general gravitational topological term is
\begin{equation}\frac{1}{2} w_2(TX) \cup (\alpha(A) + \beta(B)),\end{equation}
where $\alpha \in H^2(B\Pi_1,\bZ_2)$, $\beta: \Pi_2 \to \bZ_2$. However, by the Wu formulas
\begin{equation}\label{eqnwuformula}
\begin{gathered}
w_3(TX) \cup a = \frac{da}{2} \cup \frac{da}{2} = 0 \mod 2\\
w_2(TX) \cup b = b \cup b \mod 2,
\end{gathered}
\end{equation}
(all equations in $H^4(X,\bZ_2)$ where $X$ is an oriented 4-manifold) this is the same as
\begin{equation}\frac{1}{2} \alpha(A) \cup \alpha(A) + \frac{1}{2} \beta(B) \cup \beta(B),\end{equation}
so the gravitational terms are already captured by the ones we discussed.

There can also be coupling to gravity through the cocycle equations for $A$ and $B$. In general we have
\begin{equation}\label{eqngravcocycletwist}
    \begin{gathered}
    dA = f w_2(TX) \\
    d_\alpha B = \beta(A) + s w_3(TX),
    \end{gathered}
\end{equation}
where $f \in \Pi_1$ and $s \in \Pi_2$ are either the identity or an element of order 2. The interpretation of these modifications is that $A$ Wilson lines which pair with $f$ and $B$ Wilson surfaces that pair with $s$ are \emph{fermionic} in the sense of \cite{thorngren2015framed}.

Let us argue that $f$ is in the center of $G$. Indeed, under a change of the vertex ordering of $X$, $w_2(TX)$ may shift by an arbitrary exact cocycle
\begin{equation}\label{eqn2wtransf}
    w_2(TX) \mapsto w_2(TX) + dh,
\end{equation}
$h \in C^1(X,\bZ_2)$ \cite{thesis}. This should be thought of as a large gravitational gauge transformation, $h$ being valued in $\pi_1 SO(D)$ \cite{thorngren2015framed}. To preserve the cocycle conditions \eqref{eqngravcocycletwist}, $A$ must transform as well. If $f$ is in the center of $G$, then the action is
\begin{equation}A(ij) \mapsto A(ij) f^{h(ij)}.\end{equation}
However, if $f$ is not in the center, then there is no simple formula for this transformation, and worse, $\Pi_1$ gauge transformations would have to act on the right hand side of \eqref{eqngravcocycletwist}, so $f$ would not be gauge invariant.

For similar reasons, $f$ must also act trivially on the 2-form component $B$ by $\alpha$. Otherwise, $B$ would have to transform under the gauge transformation \eqref{eqn2wtransf} to preserve \eqref{eqngravcocycletwist}. Finally, we also need
\begin{equation}\beta(A + h) - \beta(A)\end{equation}
to be exact, which implies $\beta$ is the pullback of a 3-cocycle for the group $G/\bZ_2^f$.

All this implies that the $\bZ_2^f$ component of $A$ may be dualized to a 2-form $B'$ with topological term $\frac{1}{2} B' \cup w_2(TX)$, which we noted above is equivalent to $\frac{1}{2} (B')^2$. Thus, without loss of generality we may assume $f = 1$.

We must keep $s$ possibly nontrivial, however. When we dualize $B$ to a 1-form $\lambda$ (in the absence of a pure 2-form topological term), analogous to \eqref{eqnpostclasstopterm} we find the modified topological term
\begin{equation}-\int_X \langle \beta(A) + sw_3(TX), \lambda \rangle,\end{equation}
of which the mixed gravitational/$\lambda$ part may be reduced to a pure $\lambda$ term using the Wu formula \eqref{eqnwuformula}.

On the other hand, for gauge invariance of the pure 2-form topological term, we must have $s \in K_q$, so we may restrict $s$ to order 2 elements in $K_q$ such that $q(s) = -1$. We will see it reappear in the canonical form below, where it is reduced to a single $\bZ_2$ invariant $s = 0$ or 1, such that in the case $s = 1$, we have the pure gravitational anomaly of Section \ref{subsubsecpuregrav}. See also below.

\subsection{General Duality}\label{subsecgenduality}

So far, we have shown that the gauge theory for an arbitrary 2-group $(\Pi_1,\Pi_2,\alpha,\beta)$, with a topological term involving no pure 2-form ($j = 0$ in \eqref{eqnserre}) part, is dual to a 1-form gauge theory, whose gauge group is an extension \eqref{eqndualextension} determined by the action of $\Pi_1$ on $\Pi_2^*$ (Pontryagin dual to $\alpha$) and the mixed topological term ($j = 2$ in \eqref{eqnserre}). The Postnikov class $\beta$ defines a topological term for this 1-form gauge field which mixes the $\Pi_1$ and $\Pi_2^*$ components. All this was already derived in \cite{highersymm}.

Now we consider the most general topological terms. The pure 2-form part is known as the Pontryagin square, and is associated with a $U(1)$-valued quadratic form
\begin{equation}q: \Pi_2 \to U(1).\end{equation}
We may write it as $P_q(B) \in C^4(X,\Pi_2)$. It satisfies
\begin{equation}P_q(B + B') = P_q(B) + \langle B, B' \rangle_q + P_q(B'),\end{equation}
where we have defined a pairing on cochains using the cup product and the associated bilinear form
\begin{equation}\langle x,y \rangle_q = q(x+y) - q(x) - q(y).\end{equation}
(A quadratic form is by definition a function $q$ such that the above expression is bilinear.) We also define the associated map
\begin{equation}
\begin{gathered}
\phi_q:\Pi_2 \to \Pi_2^* \\ \phi_q(x) = \langle x,-\rangle_q
\end{gathered}
\end{equation}
and the kernel $K_q$ of this map.

The action $\alpha$ and Postnikov class $\beta$ put constraints on the quadratic form $q$. It must be $\alpha$-invariant. Further, under a 0-form gauge transformation of $B$, we have
\begin{equation}P_q(B) \mapsto P_q(B) + \langle B, \beta_1(A,g)\rangle_q + P_q(\beta_1(A,g)).\end{equation}
The second term must be cancelled by a counterterm
\begin{equation}\langle B, \zeta(A)\rangle_q,\end{equation}
where $d\zeta(A) = \beta(A)$ modulo elements of $K_q$. That is, by a redefinition of $\beta$ we can assume $\beta$ is valued in $K_q$. Meanwhile, the third term involves only $A$ and may be canceled by a gauge variation of the pure 1-form part ($j = 4$ in \eqref{eqnserre}), see \eqref{eqncanongaugeinv} below. Likewise, by studying \eqref{eqn2wtransf}, we find the fermionic Wilson string parity $s \in \Pi_2$ is in the subgroup $K_q$.

We can uniquely decompose $\Pi_2$ into products of factors $\bZ_{p^{n}}$ for primes $p$. There is an important subtley involving $p = 2$. Indeed, the group $\bZ_2$ has four $U(1)$-valued quadratic forms determined by the value of the generator $x$ (the value of the identity is zero):
\begin{equation}q(x) = k/4, \qquad k \in \bZ_4.\end{equation}
For $k = \pm 1$, we find the associated bilinear form is the unique nondegenerate one, while for $k = 2$, although the quadratic form is nontrivial, it is associated with the trivial quadratic form. Let us therefore define the subgroup $K_q^0 < K_q$ of elements for which $q$ is identically zero. We find
\begin{equation}K_q/K_q^0 = \bZ_2^r,\end{equation}
where $r$ is the number of $\bZ_{2^n}$ factors in $K_q$. For odd order elements, being in the kernel of $q$ is the same as being in the kernel of $\phi_q$.

An important caveat is that while the Postnikov class is valued in $K_q$, we cannot guarantee it to be valued in $K_q^0$. This modifies the cocycle conditions for the mixed and pure 1-form parts of the twist. We return to this point below.

We consider the sequence
\begin{equation}K_q \to \Pi_2 \to \Pi_2/K_q =: D_q.\end{equation}
($D_q$ is known as the discriminant group of $q$ \cite{KapustinSaulina,conway}). There is an extension class $\theta \in H^3(B^2 D_q,K_q)$ associated with this sequence. We can express $B$ as a pair $B_0 \in C^2(X,K_q)$, $B_q \in C^2(X,D_q)$ satisfying
\begin{equation}\label{eqnsplitcocycle}
    \begin{gathered}
    dB_q = 0\\
dB_0 = \theta(B_q) + \beta(A) + s w_3(TX).
    \end{gathered}
\end{equation}
(Recall the Postnikov class and $s$ are valued in $K_q$.) By definition, $B_0$ only appears in the action as mixed terms plus
\begin{equation}\frac{1}{2} f_j(B_0) \cup f_j(B_0),\end{equation}
where $f_j$ are the components of the quotient map $K_q \to K_q/K_q^0 = \bZ_2^r$ in the basis associated with the cyclic decomposition of $K_q$. By changing the basis to $f = f_1+ \cdots + f_r$, $f_1 + f_2$, $f_2 + f_3$, ..., $f_{r-1} + f_r$, we find we can reduce the quadratic piece to a single term
\begin{equation}\frac{1}{2} f(B_0) \cup f(B_0),\end{equation}
where $f:K_q \to \bZ_2$ is the sum of the $f_j$. The reason this works is that, despite its appearance, this quadratic term is actually a \emph{linear} function of $B_0$. In particular, if we introduce the 2nd Stiefel-Whitney class $w_2(TX)$, then by the Wu formula \eqref{eqnwuformula}
\begin{equation}\frac{1}{2} f(B_0) \cup f(B_0) = \frac{1}{2} w_2(TX) \cup f(B_0).\end{equation}

With the substitution of the quadratic term for this gravitational term, $B_0$ only occurs in linear terms in the action. It is therefore safe to dualize it to a 1-form gauge field $\lambda \in C^1(X,K_q^*)$, where $K_q^* = {\rm Hom}(K_q,U(1))$. As we have discussed, the mixed topological terms for $B_0$ and $A$ will lead to a nontrivial group extension of $\Pi_1$ by $K_q^*$ by modifying the cocycle constraint for $\lambda$. The gravitational term further modifies this constraint to
\begin{equation}d\lambda = \frac{f}{2} w_2(TX) + \eta(A),\end{equation}
where $f: K_q \to \bZ_2$ is regarded as an order 2 element of $K_q^*$ and $\eta \in H^2(B\Pi_1,K_q^*)$ comes from the mixed term $\langle B_0,\eta(A)\rangle$ and represents the group extension of $\Pi_1$ by $K_q^*$. The first term has the interpretation that $\lambda$ Wilson lines which pair nontrivially with $f$ must be treated with a framing, which turns them into fermions \cite{thorngren2015framed}. $f \in K_q^*$ may be regarded as an emergent fermion parity.

The $B_0$ cocycle constraint \eqref{eqnsplitcocycle} becomes the topological term
\begin{equation}-\int_X \langle \theta(B_q),\lambda\rangle + \langle \beta(A),\lambda\rangle + \langle s w_2(TX),\lambda\rangle,\end{equation}
the first term of which can be placed into the form of a mixed topological term between $B_q$ and $\lambda$, the second of which couples $\lambda$ and $A$, and the third term can be written as a pure-$\lambda$ term using \eqref{eqnwuformula}. By construction, the induced quadratic form on $B_q$ is \emph{nondegenerate}, which means that by completing the square (shifting the sum variable for $B_q$), we can eliminate any mixed couplings between $B_q$ and $\lambda$ or $A$. After doing this, $B_q$ is completely decoupled from the other degrees of freedom, either by cocycle constraints or topological terms. We can therefore perform the partition sum over $B_q$. Amazingly, since the induced quadratic form is \emph{nondegenerate}, this simply introduces an \emph{invertible} factor, which only depends on the signature of $X$ \cite{Taylor2001GaussSI,discretetheta}:
\begin{equation}\frac{1}{N}\sum_{B_q \in H^2(X,D_q)} e^{i\int_X P_q(B_q)} = e^{2 \pi i \sigma(X) \sigma(q)/8},\end{equation}
where $\sigma(X)$ is the signature of $X$, $\sigma(q)$ is the signature of $q$, and $N = \sqrt{|H^2(X,D_q)|}$. These invertible pieces do not contribute to the anomaly, so we discard them in Section \ref{secanom}.

Thus, we have finally reduced the general 2-group theory all the way to a theory of a (possibly fermionic) 1-form gauge field, up to an invertible piece depending only on the signature of spacetime (a gravitational theta angle). In the fermionic case, by dualizing the $\bZ_2$ subgroup generated by the emergent fermion parity $f \in K_q^*$ (cf. Section \ref{subsecgravcoupling}), we obtain a convenient canonical form for the fermionic 2-group gauge theory:
\begin{equation}\label{eqncanonform}S = \omega(a) +\frac{1}{2}(\gamma(a) + b) \cup b,
\end{equation}
where $a \in C^1(X,G)$, $b \in C^2(X,\bZ_2)$ satisfy
\begin{equation}
    \begin{gathered}
    da = 0,\\
db = \beta(a) + sw_3(YX),
    \end{gathered}
\end{equation}
where $\beta \in H^3(BG,\bZ_2)$ is the Postnikov class and $s \in \bZ_2$ indicates whether the $b$ Wilson string is fermionic (we have reintroduced it from Section \ref{subsecgravcoupling}). The solution of the gauge invariance conditions for $\omega$ and $\gamma$ proceed as in \cite{anomaliesvarious}. We find that after adding the counterterm
\begin{equation}\delta S = \frac{1}{2} \beta(a) \cup_1 b,\end{equation}
the gauge invariance conditions for the topological terms are simplified to equations involving only cochains on $BG$:
\begin{equation}\label{eqncanongaugeinv}
    \begin{gathered}
    d\gamma(a) = 0 \\
 d\omega(a) = \frac{1}{2}\gamma(a) \cup \beta(a) + \frac{1}{2} \beta(a) \cup_1 \beta(a)
    \end{gathered}
\end{equation}
where $\cup_1$ is one of the $\cup_i$ products of Steenrod \cite{steenrod}. See Appendix B.1 of \cite{seibergkapustin} or \cite{thesis} (which also has a geometric interpretation of this product) for a review. We note that even after solving this equation, we find if $s \neq 0$, then we have the gravitational anomaly
\begin{equation}\frac{1}{2} w_2(TX) \cup w_3(TX)\end{equation}
identified in \cite{thorngren2015framed} as well as
\begin{equation}\frac{1}{2} \gamma(a) \cup w_3(TX) = \frac{1}{2} \frac{d\gamma(a)}{2} \cup w_2(TX),\end{equation}
which can be cancelled by a counter-term in view of the Wu formula
\begin{equation}
    \frac{1}{2} \gamma(a) \cup w_3(TX) = \frac{1}{2} Sq^3 \gamma(a) = 0,
\end{equation}
which holds in the cohomology of any 5-manifold.

\section{Discussion}

The data defining the action \eqref{eqncanonform} matches the data of the canonical boundary condition in \cite{Lan_2019} for their so-called ``EF1" topological orders, by taking $G_b = G$, $e_2 = \gamma$ in their notation (compare \eqref{eqncanongaugeinv} with their Eq. (5)). In their work, $\gamma \in H^2(BG,\bZ_2)$ is interpreted as an extension class for how $G_b$ is extended by the emergent fermion parity $\bZ_2^f$, which we see is the result of dualizing $B$ (at the cost of introducing explicit dependence on the 2nd Stiefel-Whitney class).

The other class ``EF2" of topological order in \cite{Lan_2019} appears to be a more general sort of 3+1D TQFT. I believe they are the same as those constructed in \cite{Cui_2019}, obtained by $G$-extension of a certain fusion 2-category obtained by the Ising braided fusion category. This TQFT describes fermionic quasiparticles as well as quasistrings which behave like Kitaev wires \cite{higherbos}.


We computed anomalies only for gauge theories and EF1 topological order. If one wants to computationally exclude all known 3+1D topological orders given an anomaly, one would also need to know how to compute the anomalies of the EF2 theories. This is an interesting problem, which seems to require new techniques beyond what we have used above (although some anomaly calculations of them were performed in \cite{higherbos}). Because these EF2 theories are a kind of $\bZ_2$ extension of EF1 theories, it seems reasonable that if an EF1 theory or gauge theory cannot realize an anomaly of odd order, such as in Table I, then neither can an EF2 theory. Indeed, the symmetry fractionalization classes of the basic EF2 theory studied in \cite{higherbos} was all 2-torsion.

\section*{Acknowledgements}

I would like to thank Dominic Else for discussions which motivated me to perform these calculations, as well as Ben McMillan, Alex Takeda, and Shing-Tung Yau for topology discussions. I am grateful to Anton Kapustin for many stimulating collaborations on related projects.

\medskip

\bibliography{main}

\appendix
\onecolumngrid
\section{Basic Concepts}

This material has appeared in various places. We include it so our conventions are understood. In the author's thesis \cite{thesis}, the reader may find a detailed introduction to the subject.

\subsection{Discrete Gauge Fields}

Let $X$ be a triangulated space with ordered vertices and $G$ be a possibly nonabelian group. A 1-form $G$ gauge field $A$ on $X$ is a collection of group elements $A(ij) \in G$ for every edge $(ij)$ (our convention is to arrange the vertices so that $i<j$ in the vertex ordering), in $X$ satisfying
\begin{equation}A(ij)A(jk) = A(ik)\end{equation}
\begin{equation}
    (dA)(ijk) := A(ij)A(jk)A(ik)^{-1} = 1
\end{equation}
for every triangle $(ijk)$. We indicate the set of these objects as $Z^1(X,G)$. For $G$ abelian they form an abelian group but for $G$ nonabelian there is no group structure. Gauge transformations are parametrized by collections of group elements $g(i) \in G$ for each vertex and act on $A$ by
\begin{equation}A \mapsto A^g := g(i)A(ij)g(j)^{-1}.\end{equation}

More generally for abelian $G$ we define a \emph{$G$-valued $k$-cochain} $A \in C^k(X,G)$ as a collection of group elements
\begin{equation}A(i_0 \cdots i_k) \in G\end{equation}
For every $k$-simplex spanned by the vertices $(i_0 \cdots i_{k+1})$ (which are by convention ordered as $i_0 < \cdots < i_{k+1}$ in the vertex ordering). We define the differential $dA \in C^{k+1}(X,G)$ by
\begin{equation}\label{eqndifferential}(dA)(i_0 \cdots i_{k+1}) = \sum_{l = 0}^{k+1} (-1)^l A(i_0 \cdots \hat i_l \cdots i_{k+1}),\end{equation}
where the hat indicates that we have dropped $i_l$ from the list. The sum is over all the boundary $k$-simplices of the $k+1$-simplex $(i_0 \cdots i_{k+1})$ and the sign comes from whether the induced boundary orientation matches the orientation from the vertex ordering. A \emph{$G$-valued $k$-cocycle} is a cochain satisfying the cocycle condition
\begin{equation}dA = 0.\end{equation}
We denote the group of these cocycles as $Z^k(X,G)$.

Observe that for $k = 1$ this reduces to the definition above for abelian $G$.\footnote{There is no known way to define nonabelian $k$-cocycles for $k > 1$ since we don't know how to order the terms in \eqref{eqndifferential}.} This motivates the definition of a \emph{$k$-form $G$ gauge field} as a $G$-valued $k$-cocycle. Gauge transformations act on $A$ by shifts
\begin{equation}A \mapsto A + d\lambda\end{equation}
where $\lambda \in C^{k-1}(X,G)$. The quotient of the cocycles by the gauge transformations is $H^k(X,G)$.

Now let $R$ be a ring, $\alpha \in C^j(X,R)$, $\beta \in C^k(X,R)$. We define the \emph{cup product} $\alpha \cup \beta \in C^{j+k}(X,R)$ by
\begin{equation}(\alpha \cup \beta)(i_0 \cdots i_{j+k}) = \alpha(i_0 \cdots i_j)\beta(i_j \cdots i_{j+k}).\end{equation}
The vertex ordering is important here, but if $\alpha$ and $\beta$ are cocycles, it turns out
\begin{equation}\alpha \cup \beta = (-1)^{jk} \beta \cup \alpha + d(\cdots).\end{equation}
The counterterms in $\cdots$ define the $\cup_1$ product \cite{steenrod}.

\subsection{2-Group Gauge Fields}

We review some basic notions from \cite{highersymm}. See also \cite{Baez_2010,crane1993categorical,Crane_1997}.

A 2-group $(\Pi_1,\Pi_2,\alpha,\beta)$ is specified by a (possibly nonabelian) finite group $\Pi_1$, an abelian finite group $\Pi_2$, and action $\alpha:\Pi_1 \to {\rm Aut}(\Pi_2)$, and a ``Postnikov class" $\beta \in H^3(B\Pi_1,\Pi_2)$, which can be thought of as a natural map (not necessarily a homomorphism)
\begin{equation}\beta:C^1(X,\Pi_1) \to C^3(X,\Pi_2)\end{equation}
satisfying
\begin{equation}d\beta(A) = 0 \qquad {\rm when\ } dA = 0\end{equation}
and
\begin{equation}\beta(A^g) = \beta(A) + d\beta_1(A,g),\end{equation}
for some function
\begin{equation}\beta_1:C^1(X,\Pi_1) \times C^0(X,\Pi_1) \to C^2(X,\Pi_2)\end{equation}
known as the first descendant of $\beta$.

A $(\Pi_1,\Pi_2,\alpha,\beta)$-valued gauge field on $X$ is a pair
\begin{equation}A \in C^1(X,\Pi_1)\end{equation}
\begin{equation}B \in C^2(X,\Pi_2)\end{equation}
satisfying the cocycle condition
\begin{equation}dA = 0\end{equation}
\begin{equation}d_\alpha B = \beta(a),\end{equation}
where $d_\alpha B$ is the twisted differential
\begin{equation}(d_\alpha B)(ijkl) = (dB)(ijkl) - 2 \alpha(A(ij)) \cdot B(jkl),\end{equation}
where in the second term we use the action $\alpha$ of $\Pi_1$ on $\Pi_2$. A gauge transformation is parametrized by a pair
\begin{equation}g \in C^0(X,\Pi_1)\end{equation}
\begin{equation}\lambda \in C^1(X,\Pi_2)\end{equation}
and acts by
\begin{equation}A \mapsto A^g\end{equation}
\begin{equation}B(ijk) \mapsto \alpha(g(i)) \cdot B(ijk) + (d\lambda)(ijk) + \beta_1(A,g)(ijk),\end{equation}
where $\beta_1$ is the first descendant of $\beta$, defined implicitly above. This extra term is needed to preserve the cocycle equation for $B$.

\section{Computing the Classification}\label{appcomptheclass}

The classifications \eqref{eqnclassanom} and \eqref{eqnclassanomtrs} can be computed using the following standard facts (see eg. \cite{cohomologyoperations}):
\begin{enumerate}
    \item $H^j(B^2 K_1, \bZ) = 0$ if $j = 1, 2$ or 4 (although the latter can be nonzero if $K_1$ has infinitely many components).
    \item $H^3(B^2 K_1,\bZ) = F \times \bZ^r$, where $K_1 = F \times U(1)^r$ and $F$ is finite.
    \item $H^5(B^2 K_1,\bZ) = H^4(B^2 F, U(1))$, the group of $U(1)$-valued quadratic forms on $F$.
    \item $H^6(B^2 K_1,\bZ) = H^5(B^2 F, U(1)) \times \bZ^{r(r-1)/2}$, the latter being the Chern-Simons terms for the continuous part of $K_1$.
    \item $H^j(B^2 K_1, \bZ_2) = 0$ if $j = 1, 3$.
    \item $H^2(B^2 K_1, \bZ_2) = {\rm Hom}(K_1,\bZ_2)$.
    \item $H^4(B^2 K_1, \bZ_2) = {\rm Hom}(K_1,\bZ_2)$ through $B \cup B = Sq^2 B$.
    \item $H^5(B^2 K_1, \bZ_2) = {\rm Hom}(K_1,\bZ_2)^2$ through $B Sq^1 B$ and $Sq^2 Sq^1 B$. These are both equivalent by a Wu formula to the $\theta = \pi$ term $\frac{1}{2} \left(\frac{dB}{2}\right)^2$.
\end{enumerate}
Thus, the $k = -1$ part of \eqref{eqnclassanom} is only nonzero for $j = 0$, 3, 5, or 6:
\begin{enumerate}
    \item $H^6(BK_0,\bZ)$, the pure-0-form terms.
    \item $H^3(BK_0,\bZ)$, central extensions of $K_0$.
    \item $H^1(BK_0,H^4(B^2 F,U(1))) = {\rm Hom}(K_0, H^4(B^2 F,U(1)))$.
    \item $H^6(B^2 K_1, \bZ)$, the pure-1-form terms.
\end{enumerate}
And the $k = 3$ part of \eqref{eqnclassanom} is nonzero only for $j = 0$:
\begin{enumerate}
    \item $H^2(BK_0,\bZ) = {\rm Hom}(K_0, U(1))$, which labels mixed gravitational Chern-Simons terms of type $A \wedge R \wedge R$.
\end{enumerate}
Also, the $k = 1$ part of \eqref{eqnclassanomtrs} is only nonzero for $j = 0,$ 2, or 4:
\begin{enumerate}
    \item $H^4(BK_0, \bZ_2)$, terms of the form $\frac{1}{2} w_1(TX) \zeta(A)$. Those $\zeta$ with integer lifts are killed by the Wu relations.
    \item $H^2(BK_0, H^2(BK_1, \bZ_2))$, terms of the form $\frac{1}{2} w_1(TX) c(A) f(B)$, with $c \in H^2(BK_0,\bZ_2)$ and $f: K_1 \to \bZ_2$.
    \item $H^4(BK_1,\bZ_2)$, terms of the form $\frac{1}{2} w_1(TX) P(B)$. Those $P$ with integer lifts are killed by the Wu relations.
\end{enumerate}

The Wu formulas apply to \eqref{eqnclassanom} by simply eliminating the $k = 2$ piece. In \eqref{eqnclassanom}, as well as killing certain terms above, one $k = 2$ piece remains, for anomalies of the form
\[\frac{1}{2} w_1(TX)^2 \rho(A,B),\]
with $\rho$ a $\bZ_2$ 3-cocycle. The classification of $\rho$ splits into three cases, $j = 0, 2$, or 3:
\begin{enumerate}
    \item $H^3(BK_0,\bZ_2)$, a pure 0-form and TRS anomaly.
    \item $H^1(BK_0, H^2(BK_1, \bZ_2)) = {\rm Hom}(K_0 \times K_1, \bZ_2)$, terms of the form $\frac{1}{2} w_1 f(A,B)$, where $f$ is a pairing $K_0 \times K_1 \to \bZ_2$. We can write any such map as a product of maps into $\bZ_2$, since $f$ must factor through the finite part of the abelianization of $K_0$ as well as the finite part of $K_1$, and the product of finite abelian groups is also the coproduct.
    \item $H^3(BK_1,\bZ_2) = \bZ_2^r \times {\rm Hom}(F,\bZ_2)$.
\end{enumerate}


\end{document}